# Pressure-constant Monte Carlo simulation of phase I of solid $CO_2$ up to 10 GPa at $T$ = 200 K using Kihara potential model

Koji Kobashi

Former Research Assistant, Physics Department, Colorado State University, Fort Collins, CO, USA,

and

Former Senior Researcher, Kobe Steel, Ltd., Japan

*Abstract*

This article is a continuation of the past three papers, arXiv:1711.04976 (2017), arXiv:1809.04291 (2018), and arXiv:2006.09673v2 (2020), in which configurations of the molecules around a vacancy in solid $CO_2$ with the *Pa3* structure (phase I) were calculated by the Monte Carlo (MC) simulation technique at $T \leqq 200$ K and at a *nominal* pressure of $P$ = 1 atm using lattice constants determined in reference to experimental data. For the intermolecular potential, the Kihara model of a rod-shape core with zero diameter was used. For theoretical consistency, however, the lattice constant should be determined by a pressure-constant MC simulation, *i.e.* by the *NPT* simulation. It was anticipated that the *NPT* simulation, successful for monoatomic molecular fluids, might not work straightforwardly for solid $CO_2$ because of the non-spherical molecular shape and interactions. In fact, it was found in the present work that the protocol of the standard *NPT* simulation had to be modified for solid $CO_2$ to achieve an equilibrium lattice constant, given an external pressure, within practical computer time and capacity. The modified protocol was used to calculate the *P-V* relation at 1 bar and from 2 to 10 GPa at which a structural phase transition is known to take place. The temperature was fixed at 200 K. As a result, the calculated *P-V* relation was similar to those in previous theoretical works, the molar volume being roughly 0.5 % smaller than experimental data.

1. Introduction

The present work is a continuation of the precedent three articles:[1-3] the first two articles were concerned with molecular configurations surrounding a single vacancy in solid $CO_2$ with the *Pa3* structure at temperatures $T$ = 0, 100, and 200 K: the third article was concerned with a larger vacancy in which a central molecule and its first and second neighbor molecules were removed to make a cubic vacancy, and molecular configurations forming the vacancy wall were studied. In those calculations, the Monte Caro (MC) simulation technique[4] and a Kihara core potential model were used.[5] The basic cell was cubic that contained $N$ = 2048 molecules, eight primitive unit cells on each side, to which the periodic boundary condition was applied. The lattice constants $a$ used at different temperatures were assumed in reference to experiments[6] as follows: $a$ = 5.53725 + 4.679 $\times 10^{-6}$ $T^2$ in units of Å (1 Å = 0.1 nm). Using experimental or assumed lattice constants is convenient for such studies as molecular librations in linear molecular solids with empirical intermolecular potentials[7] because the temperature dependence is included in $a$. To theoretically determine the lattice constant under pressure $P$, it is necessary to undertake the $P$-constant simulation, or the *NPT* simulation.[4] It is anticipated, however, that the standard protocol of the *NPT* simulation, which is successful for monoatomic molecular fluids, might not work straightforwardly for simple linear molecular systems such as solid $CO_2$. The major purpose of the present work is to examine and modify the standard protocol of the *NPT* simulation and then to investigate to what extent the modified protocol is feasible to evaluate the *P-V* relation for solid $CO_2$.

Before moving to the detail of the present work, it would be worthwhile to describe a brief history of the research on simple molecular crystals and solid $CO_2$ as well. Researches of simple molecular solids were intensively carried out in the 1970s - 80s as the low temperature technology was developed to condense gases such as $N_2$ and $CO_2$. For theoretical studies, empirical pair potentials were used: the most widely-used potential model was to assume a Lennard-Jones (LJ) potential between atomic centers of difference molecules, and then sum them up. For the electrostatic multipole-multipole potential, fractional charges were placed on each atom, or a point multipole was placed in the center of each molecule. In the Kihara potential model of $CO_2$, a rod-shape core with a length of $l$ = 2.21 Å and zero diameter is placed inside the molecule, and a LJ potential is assumed between the shortest distance $\rho$ of the cores associated with different molecules. In addition to the LJ potential, a point electrostatic quadrupole is placed in the center of each molecule. It was unfortunate that MC and Molecular Dynamics (MD) simulations for simple linear molecules were not done intensively in that period because of insufficient processing speed and capacity of computers. Moreover, research of molecular solids and fluids was so fundamental that there was no practical use of the results. For those reasons, research of molecular solids and fluids slowed down in the 1990s. However, as the processing speed, the storage capacity, and the computational cost were improved significantly in the 2000s, research of both monoatomic and simple linear molecules became possible based on MC and MD simulations. It is notable that intermolecular potentials can be theoretically determined owing to the development of quantum electronic theories of molecules: such theories are particularly suitable for studying molecular crystals under extremely high pressure at which the atoms of a molecule start to form chemical bonds with the atoms associated with adjacent molecules. Furthermore, research results on the studies of solid and fluid properties $CO_2$ become of practical use recently: for instance, $CO_2$, as a global warming gas, is planned to be confined deep in the bottom of an ocean under high pressure, and precise properties of solid and fluid $CO_2$ are required for developing the technology. It is expected that equations of states of simple molecules such as $H_2$, $NH_3$, and $CO_2$ will be more intensively studied with regards to social issues of energy and environments in the years to come.

Now coming back to the present work, Fig. 1 shows the phase diagram of $CO_2$ under high pressure. In the present work, the *P-V* relation was computed by the *NPT* simulation for phase I with the space group *Pa3* at 200 K up to 10 GPa at which a structural phase transition to phase III is known to take place.[8-13] In phase I, studies of solid $CO_2$ using empirical potential models have an advantage that the computational time and cost are small enough so that the *NPT* simulations could be done even on regular desktop computers.

In the following, the computational protocols for the *NPT* simulation will be described in Sec. 2, results and discussion in Sec. 3, and finally, a conclusion in Sec. 4. As a result of the present work, a new protocol for the *NPT* simulation, suitable for solid $CO_2$, was identified although it needs to be further elaborated.

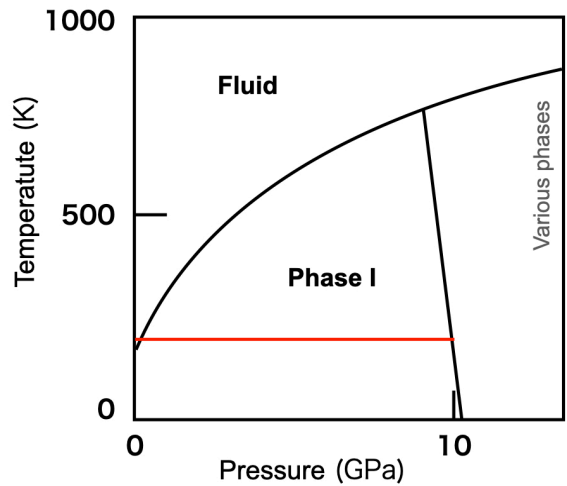

Fig. 1. *P-T* phase diagram of $CO_2$.

The *P-V* relation at 200 K in phase I is computed along the red line in the present work.

2. Computational procedure

    As described above and in Refs. 1 and 5 as well, the Kihara core potential with a 9-6 LJ potential and a point electrostatic quadrupole-quadrupole (EQQ) interaction was assumed between molecules for energy calculations of solid $CO_2$. The maximum interaction range of the 9-6 LJ potential between the cores was set to be 10 Å, while the maximum interaction range of both the van der Waals potential, in which the repulsive part of the 9-6 LJ potential was removed, between molecular centers and the EQQ potential were set to be 15 Å. One could refer to Ref. 5 for more details of the intermolecular potential. The theory and the protocol of the *NPT* simulation are described in the textbook[4] of Frenkel and Smit, and a schematic of the standard protocol is shown in Fig. 2.

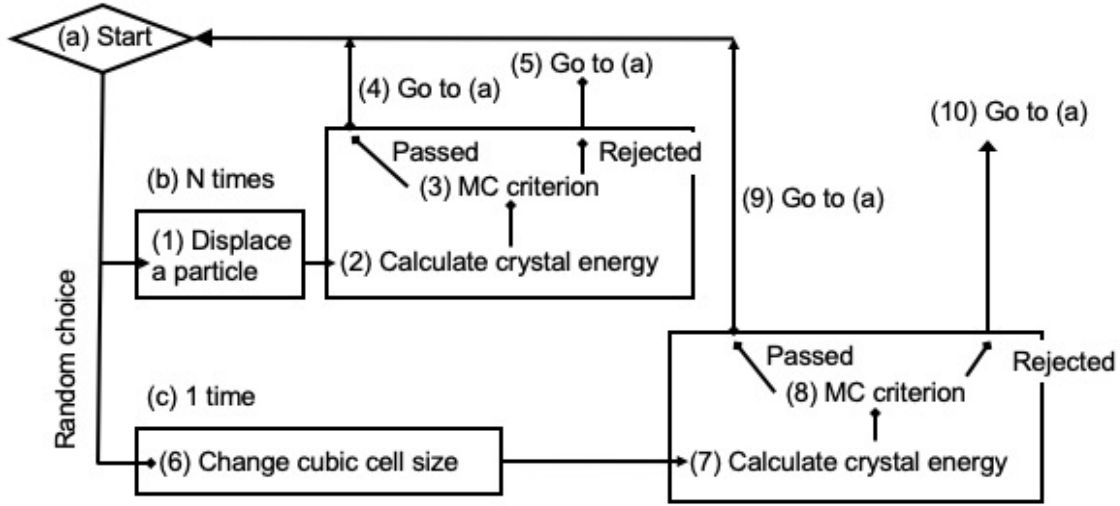

Fig. 2. Standard protocol of *NPT* simulation for monoatomic (particle) systems.[4]

The calculations begin from (a) *Start*, and then randomly choose either (1) *Displace a particle* or (6) *Change cubic cell size* with the frequency ratio of (1) : (6) = Number of molecules $N$ (= 2048) : 1. Namely, a random displacement of a particle and a random change in the basic cubic cell size are equally treated so that (6) *Change cubic cell size* is done only 1/(2048+1) on average. In case that (1) *Displace a particle* is chosen, the crystal energy is calculated after a randomly chosen particle is displaced. If the displacement is accepted by the MC criterion,[4] then the computational procedure comes back to (a) *Start* with the new configuration. Else, the crystal structure comes back to (a) *Start* without any change. On the other hand, in case that (6) *Change cubic cell size* is chosen, the basic cubic cell size is randomly and uniformly changed, and the crystal energy is calculated. Again, if the new cell size is accepted by a criterion for the volume change,[4] the new cell size is used for the subsequent calculations at (a) *Start*. Else, the cell size is unchanged. In the long run, MC attempts are done uniformly for both the configurations of 2048 molecules and the cell size. It was however found that if this protocol is used, (6) *Change cubic cell size* was successful only once or twice among the entire rounds of 1,000,000. Thus, this protocol was not effective to determine the basic cubic cell size, and hence modified as shown in Fig. 3.

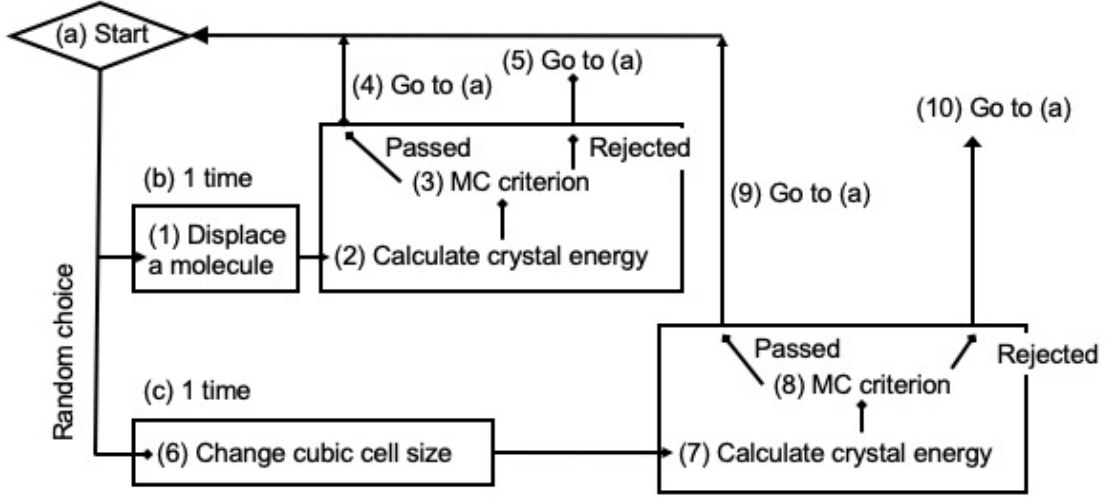

Fig. 3. Protocol used in the present work.

In the protocol of Fig. 3, the probability of attempts to (1) *Displace a molecule* and to (2) *Change cubic cell size* was set equal so that among 1,000,000 attempts, the half were done to change the basic cubic cell size. The random volume change of the cell was done according to the following equation:[4]

$$\ln(V_n) = \ln(V_0) + (2\,\text{rand}() - 1)\,V_{max}, \qquad (1)$$

where $V_0$ and $V_n$ are an initial and a new cell volumes, respectively, rand() is a random number, and $V_{max}$ is a parameter determined to be $V_{max} = 20$. Therefore, $\ln(V_n)$ is randomly determined within a range:

$$\ln(V_0) - V_{max} < \ln(V_n) < \ln(V_0) + V_{max}. \qquad (2)$$

It should be noted that even in this protocol, where the number of attempts to change the basic cubic cell size was as many as about 500,000, only about 35 cases were accepted. Even so, the cell size converged only after three rounds of computations. In case that (1) *Displace a molecule* was selected, the same protocol as in Fig. 2 was performed. In addition to the protocol of Fig. 3, the standard procedure of MC simulations, used in the past three papers,[1-3] were performed to equilibrate the molecular configurations in the new cell. It should be noted at this point that (i) the range of the random displacements for the molecular center was -0.1 Å < $\Delta x, \Delta y, \Delta z$ < +0.1 Å, and that for the molecular orientations was -30° < $\Delta\theta$, $\Delta\phi$ < +30°, or {0.1, 30} in abbreviation; and (ii) in preliminary computations for $V_{max}$ = 20, 10, 5, and 1 in Eq. (1), the numbers of successful attempts to change the cell size were 102, 209, 415, and 1866, respectively. However, as stated above, the cell size converged in the first three rounds of computations irrespective of the $V_{max}$-value. The calculation procedure and the calculated results for the case of $P$ = 10 GPa are shown in Table 1. For the notations in Table 1, see the captions under the Table.

The computations were done on regular desktop computers using gfortran, a Linux Fortran. The computation times are shown in Table 1. To achieve equilibrium configurations of the molecules in the basic cubic cell, the regular MC simulations,[1-3] shown in the 3rd row of the column denoted as "Step", have to be repeated until the crystal energy converges: however, in the present work, it was done only once in the middle of the calculations, because the major interest of the present work is to examine the protocol shown in Fig. 3, and the basic cubic cell size converged quickly irrespective of the repetition of the regular MC simulations.[1-3]

Table 1. Calculation procedure for solid $CO_2$ at 200 K and 10 GPa.

| Step | Program | Molar volume (cm³/mol) | Cell size (Å) | No. of rounds | No. of (1) in Fig. 3 | No. of (6) in Fig.3 | No. of changes in cell size | No. of molecular moves | Crystal energy (K) | Computation time (min) |
|---|---|---|---|---|---|---|---|---|---|---|
| 1 | Fig. 3 | 18.7944 | 39.9816 | 1,000,000 | 499,536 | 500,464 | 36 | 3,272 | -1,089,891.0 | 246 |
| 2 | Fig. 3 | 18.7975 | 39.9838 | 1,000,000 | 499,765 | 500,235 | 22 | 1,204 | -1,119,360.8 | 243 |
| 3 | MC | Same as above | Same as above | 100,000 | | | | 9,654 | -1,038,711.4 | 58 |
| 4 | Fig. 3 | 18.7860 | 39.9756 | 1,000,000 | 499,120 | 500,880 | 34 | 2,407 | -1,080,378.7 | 245 |

Step: computation step, Program: protocol used for the computation step, Molar volume: calculated molar volume in units of (cm³/mol), Cell size: edge length of the basic cubic cell in units of Å, No. of rounds: total number of computation cycles, No. of (1) in Fig. 3: number of choosing (1) in Fig. 3, No. of (6) in Fig. 3: number of choosing (6) in Fig. 3, No. of changes in cell size: number of changes in the basic cubic cell size in the computation step, No. of molecular moves: number of molecular displacements in the computation step, Crystal energy: total crystal energy in units of K (1 K = $1.38065 \times 10^{-23}$ J), and Computation time in units of minutes. Note that as shown in the column of Program, the protocol of Fig. 3 was repeated twice, then the standard MC simulation was performed once, and finally the protocol of Fig. 3 was done just once.

3. Results and Discussion

The calculated *P-V* relation (open circles) is shown in Fig. 4 along with experimental data (solid circles).[14] The calculated molar volume was roughly 0.5 % lower than experimental data. The difference between theory and experiment tend to be smaller as the pressure increases. This result was very similar to those in Refs. 8 and 9, in which intermolecular potential was determined by quantum mechanical calculations. Therefore, even with quite a simple computation protocol and a small number of computations in the present work, the calculated results were found to be reasonable.

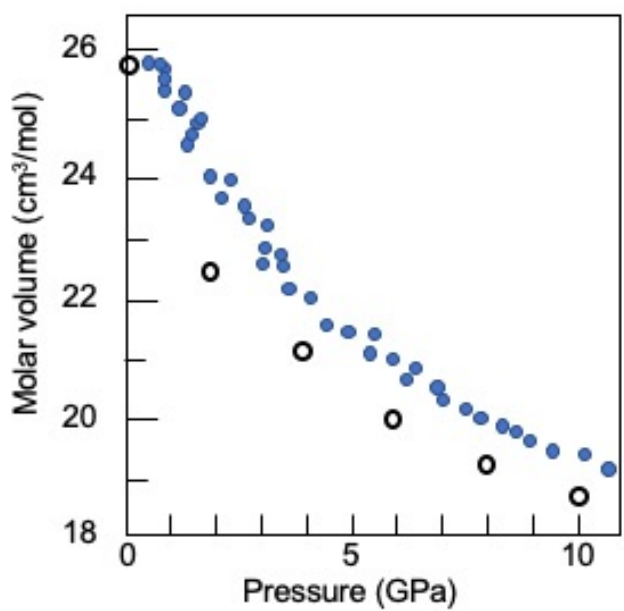

Fig. 4. *P-V* relation; open circles: calculated results, solid circles: experimental data.[14]

Figure 5 shows the configurations of the molecules at (a) 1 bar and (b) 10 GPa at *T* = 200 K, observed from the [111] direction of the *Pa3* crystal. A brown sphere and two red spheres in a molecule indicate carbon and oxygen atoms, respectively, with van der Waals radii. The three molecules in the center of the figures are located at the very end of the basic cubic cell. It is clearly seen that the molecules are more closely packed at 10 GPa than at 1 bar.

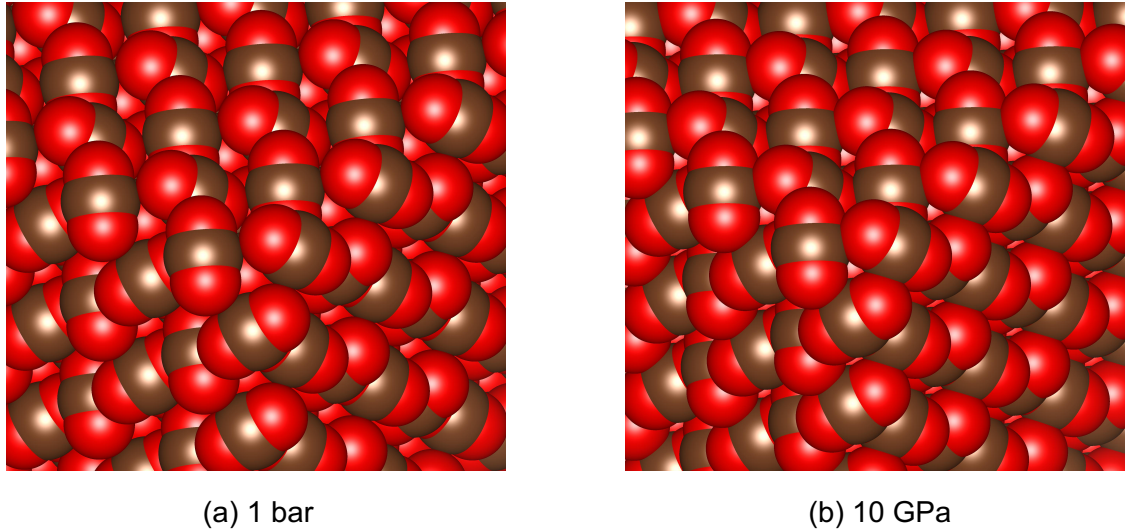

(a) 1 bar    (b) 10 GPa

Fig. 5. Crystal structure of phase I at (a) 1 bar and (b) 10 GPa

4. Conclusion

In order to perform *NPT* simulations within practical computation time and computer capacity, the standard *NPT* simulation protocol was modified so that the molecular displacement and the change in the basic cubic cell size are done with an equal probability. Consequently, theoretically reasonable *P-V* relation was obtained at $T = 200$ K and $P = 10$ GPa. Since the present work is the first attempt for the *NPT* simulation in a series of research on solid $CO_2$, further improvements of the simulation protocol will be necessary: (i) the first is the computational steps, shown in Table 1. This must be optimized in terms of both the number and the order of the protocols shown in Fig. 3 as opposed to the regular MC simulations. For instance, the number of computation steps must be increased until the crystal energy is stable; (ii) the second is that the parameter, $V_{max}$, must be optimized at the stage of the computation; (iii) the third is to further improve the protocol of Fig. 3 so that the equilibrium basic cubic cell size and molecular orientations are quickly attained; (iv) finally, the potential parameters in the Kihara model may be modified so as to get a better agreement of the *P-V* relation with experiment (see Fig. 4). In summary, the present work showed that the protocol of the *NPT* simulation for monoatomic solids and fluids did not work straightforwardly for linear molecular systems such as solid $CO_2$, and a new protocol is required to be established.

*Acknowledgement*

Figures 5 is depicted using an open software, VESTA, published in: K. Momma and F. Izumi, "*VESTA 3 for three-dimensional visualization of crystal, volumetric and morphology data*," J. Appl. Crystallogr. **44**, 1272 (2011).

*References*

1. K. Kobashi, arXiv:1711.04976 [cond-mat.mtrl-sci] (2017).

2. K. Kobashi, arXiv:1809.04291 [cond-mat.mtrl-sci] (2018).

3. K. Kobashi, arXiv:2006.09673v2 [cond-mat.mtrl-sci] (2020)

4. D. Frenkel and B. Smit, *Understanding Molecular Simulation (Second Edition)* (Academic Press/Elsevier, London, 2001).

5. K. Kobashi and T. Kihara, J. Chem. Phys. **72**, 3216 (1980).

6. W. H. Keesom and J. W. L. Köhler, Physica (The Hague), **1**, 655 (1934).

7. S. Califano, V. Schettino, and N. Neto, *Lattice Dynamics of Molecular Crystals* (Springer/Elsevier, London, 1981).

8. J. Li, O. Sode, G. A. Voth, and S. Hirata, Nature Commun. No. 2647, 4 (2013).

9. S. Hirata, K. Gilliard, X. He, J. Li, and O. Sode, Acc. Chem. Res. **47**, 2721 (2014).

10. Y. Han, J. Liu, L. Huang, X. He, and J. Li, npj Quantum Mater. 4 (2019).

11. V. M. Giordano, and F. Datchi, arXiv:0608529v1 [cond-mat.other] (2006).

12. S. A. Bonev, F. Gygi, T. Ogitsu, and G. Galli, arXiv:0305370v1 [cond-mat] (2003); *ibid*. Phys. Rev. Lett. **91**, 065501 (2003).

13. B. H. Cogollo-Olivo, S. Biswas, S. Scandolo, and J. A. Montoya, arXiv:1908.11352v2 [cond-mat.mtrl-sci] (2019); *ibid*. Phys. Rev. Lett. **124**, 095701 (2020).

14. See Refs. 8 and 9 for the sources of the experimental data.